\documentstyle[12pt]{article}
\begin{document}

\title{
Simulation of Crystal Extraction Experiments}
\author{Valery M. Biryukov\thanks{E-mail:  biryukov@mx.ihep.su  }
\\ \vspace{5mm} IHEP Protvino, 142284 Moscow Region, Russia
\\ Published in Near Beam Physics Symp. Proc.,
\\ Fermilab-Conf-98/166, pp.179-184 (1998) }
\date{Presented at Fermilab, 22-24 Sep 1997}

\maketitle

\abstract{
We discuss the simulation methods
and results for the crystal extraction experiments
performed recently at the high energy accelerators.
Possible future applications of the
crystal channeling technique are considered.
}

\section{Introduction}

Crystal extraction experiments have greatly progressed
in recent years,
spanning over two decades in energy and more than two decades
in the crystal bending angle\cite{ass,sps,e853}.
The theory of crystal extraction is essentially based on
Monte Carlo simulations, as the
extraction process includes multiple passes through the
crystal, and turns in the accelerator, of the beam particles.
Even more importantly, tracking of a particle through a bent
crystal lattice requires not only a calculation of a particle
dynamics in this nonlinear field, but also a generation of random events
of scattering on the crystal electrons and nuclei.

To track particles
through the curved crystal lattices in simulation we apply
the approach with a continuous potential
 introduced by Lindhard.
In this approach one considers collisions of the incoming
particle with the atomic strings or planes instead of with
separate atoms, if the particle is sufficiently aligned with respect to
 the crystallographic axis or plane.
 The typical step size along the crystal length in simulation
 is about 1 micron, as defined by the particle dynamics in crystal channel.
By every step the probabilities of scattering events
on electrons and nuclei are computed depending on their local densities
which are functions of coordinates.
This ensures correct orientational dependence of all the processes
in crystal material.
Further details on the simulation code may be found
in Refs.\cite{pre,pre3}.

Leaving aside the details of channeling physics, it may be useful
to mention that accelerator physicist will find many familiar things there:
\begin{itemize}
\item
Channeled particle oscillates in a transverse nonlinear field of
a crystal channel, which is the same thing as the
"{\em betatronic oscillations}" in accelerator,
but on a much different scale (the wavelength is 0.1 mm at 1 TeV
in silicon crystal).
The number of oscillations per crystal length can be
several thousand in practice.
The concepts of beam emittance, or particle action
have analogs in crystal channeling.
\item
The crystal nuclei arranged in crystallographic planes
represent the "{\em vacuum chamber walls}". Any particle approached
the nuclei is rapidly lost from channeling state.
Notice a different scale again: the "vacuum chamber" size is $\sim$2 \AA.
\item
The well-channeled particles are confined far from nuclei
(from "aperture"). They are lost then only due to scattering
on electrons. This is analog to "{\em scattering on residual gas}".
This may result in a gradual increase of the particle amplitude
or just a catastrophic loss in a single scattering event.
\item
Like the real accelerator lattice may suffer from {\em errors of
alignment},
the lattice of real crystal may have dislocations too,
causing an extra diffusion of particle amplitude
or (more likely) a catastrophic loss.
\item
Accelerators tend to use low temperature, superconducting magnets.
Interestingly, the crystals cooled to {\em cryogenic temperatures}
are more efficient, too.
\end{itemize}

\section{The SPS Experiments}

A detailed account for the crystal extraction experiments
made at the CERN SPS can be found in this volume\cite{sps}.
Before these SPS studies, the theoretical comparisons \cite{bi91}
with extraction experiments \cite{dubna,ass1}
were restricted by  analytical estimates only,
which gave the right order of magnitude.
The computer simulations considered idealized models only
and predicted the extraction efficiencies
always in the order of 90--99\% (e.g. \cite{bi91})
while real experiments handled much smaller efficiencies,
in the order of 0.01 \% \cite{dubna,ass1}.

The considered-below theoretical work has been the first and
rather detailed comparison between the
realistic calculation from the first principles (computer simulation)
and the experiment.
The simulation was performed \cite{bi78}
with parameters matching those of the SPS experiment.
Over 10$^5$ protons have been tracked
both in the crystal and in the accelerator for many subsequent
passes and turns until they were lost either at the
aperture or in interaction with crystal nuclei.

In the simulation, different
assumptions about quality of the crystal surface were applied:
one was an ideal surface, whereas the other one
assumed near-surface irregularities
(a `septum width') of a few $\mu$m due to a miscut angle
(between the Si(110) planes and the crystal face) 200\,$\mu$rad,
surface nonflatness 1\,$\mu$m,
plus 1\,$\mu$m thick amorphous layer superposed.
Two options were considered. The
{\em first}, with impact parameter below 1\,$\mu$m and
surface parameters as described above, excludes the possibility
of channeling in the first pass through the crystal.
This is compared to the {\em second} option, in which
the crystal surface is assumed perfect, i.e., with a zero septum width.

 Table~1 shows the expected extraction efficiencies
 for both options from the first simulation run
 and the measured lower limit of extraction efficiency
 as presented at the 19-th meeting on "SPS Crystal Extraction"
 \cite{meet19} held at CERN.

 \begin{table}
 \caption{SPS crystal extraction efficiencies
 from the early runs, Monte Carlo and experiment}
 \begin{tabular}{lcc}
 & & \\
 \hline
 & & \\
 Option & Monte Carlo & Experiment   \\
 & & \\
 \hline
 & & \\
 Poor surface & 15\% & lower limit \\
	   &       & of 2-3\%        \\
 Ideal surface & 40\% & only known \\
 & & \\
 \hline
 \end{tabular}
 \end{table}

Though the efficiency comparison, theory to  measurements,
was not possible at that time,
from the analysis of the simulation results
one could see that the perfect-surface simulation predicted
narrow high peaks for the angular scans (30 $\mu$rad FWHM) and
extracted-beam profiles, which have not been observed.
The imperfect-surface option, however, is approximately consistent
with the experimental observations: wide (about 200 $\mu$rad FWHM)
angular scan and sophisticated profiles of the extracted beam
(dependent on the crystal alignment).

The efficiency was measured in the SPS experiment
with that first tested crystal to be
10$\pm$1.7\%. The detailed simulations have shown that
efficiency should be a function of the vertical coordinate
of the beam w.r.t. the crystal (for its given shape),
and be from 12 to 18\% at peak, with imperfect-surface option.

The simulation studies
for a new crystal with another geometry
(``U-shaped'')
were performed prior to the measurements.
The model followed the parameters and design of this crystal,
with the same SPS setting.
Again the two options, an imperfect or perfect edge,
have been studied.

\begin{figure}[htb]
\begin{center}
\setlength{\unitlength}{.9mm}
\begin{picture}(70,100)(-6,-24)
\thicklines

\put(0,-19) {\line(1,0){65}}
\put(0,-19) {\line(0,1){90}}
\put(0,71) {\line(1,0){65}}
\put(65,-19) {\line(0,1){90}}
\multiput(0,-19)(0,3.994){23}{\line(1,0){1.2}}
\multiput(0,-19)(0,19.97){5}{\line(1,0){1.6}}
\put(-7,0.99){\makebox(2,1)[l]{5}}
\put(-7,20.6){\makebox(2,1)[l]{10}}
\put(-7,40.83){\makebox(2,1)[l]{15}}
\put(-7,60.8){\makebox(2,1)[l]{20}}
%\multiput(65,-19)(0,5.4){16}{\line(-1,0){1.2}}
%\multiput(65,-19)(0,27){4}{\line(-1,0){1.6}}
%\put(67,8){\makebox(2,1)[r]{1}}
%\put(67,35){\makebox(2,1)[r]{2}}
%\put(67,62.2){\makebox(2,1)[r]{3}}

\multiput(6.36,-19)(5.128,0){10}{\line(0,1){1}}
\multiput(6.36,-19)(25.64,0){3}{\line(0,1){1.5}}
\put(30,-16){\makebox(2,1)[b]{-230}}
\put(4.36,-16){\makebox(2,1)[b]{-330}}
\put(55.64,-16){\makebox(2,1)[b]{-130}}

\put(-5,73){I (rel.)}
%\put(-5,73){F(\%)}
%\put(49,73){I($\times 10^5$)}
\put(12,-24){ crystal angle ($\mu$rad)}

\put( 3.69, 4.8){ $\otimes$}
\put(16.51,13.3){ $\otimes$}
\put(24.20,21.7){ $\otimes$}
\put(29.33,57.7){ $\otimes$}
\put(34.46,19.7){ $\otimes$}
\put(42.15,14.2){ $\otimes$}
\put(54.97,12.3){ $\otimes$}

\put(23,12){70 $\mu$rad}
\put(30,19){\vector(-1,0){9.2}}
\put(30,19){\vector(1,0){8.75}}

\put(11,8){ $\star$}
\put(12,8.7){ $\star$}
\put(13,10.5){ $\star$}
\put(14,10.5){ $\star$}
\put(15,11.5){ $\star$}
\put(16,16.4){ $\star$}
\put(17,18){ $\star$}
\put(18,15.4){ $\star$}
\put(19,19.5){ $\star$}
\put(20,25.5){ $\star$}
\put(21,23.7){ $\star$}
\put(22,26.3){ $\star$}
\put(23,31.5){ $\star$}
\put(24,36){ $\star$}
\put(25,38){ $\star$}
\put(26,36.3){ $\star$}
\put(27,42){ $\star$}
\put(28,43.5){ $\star$}
\put(29,47){ $\star$}
\put(30,57.7){ $\star$}
\put(31,53.3){ $\star$}
\put(32,52){ $\star$}
\put(33,42.1){ $\star$}
\put(34,33){ $\star$}
\put(35,29.7){ $\star$}
\put(36,26.8){ $\star$}
\put(37,19.9){ $\star$}
\put(38,18.5){ $\star$}
\put(39,12){ $\star$}
\put(40,8.5){ $\star$}
\put(41,10){ $\star$}
\put(42,9){ $\star$}
\put(43,8.5){ $\star$}
\put(44,10){ $\star$}
\put(45,5.5){ $\star$}
\put(46,7.4){ $\star$}

\end{picture}
\end{center}
 \caption
  {
The angular scan of extraction with a U-shaped crystal.
  Prediction ($\otimes$) and measurement ($\star$).
}
\label{u}
\end{figure}
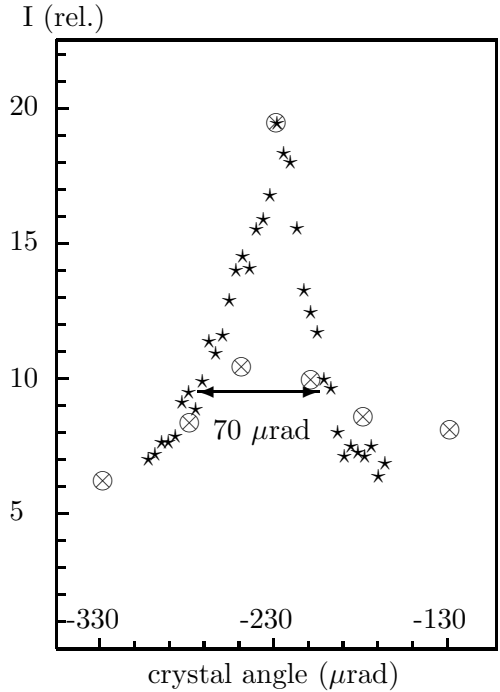

Figure~\ref{u} shows the angular scan
(as narrow as 70 $\mu$rad FWHM)
of the efficiency simulated for the U-shaped crystal
with edge imperfections; a comparison to
the measurements shows a good agreement.
The peak efficiency, 19.5$\pm$0.7\%, was expected
to be just slightly increased with the new crystal.
For an ideal crystal and a parallel incident beam,
the simulation predicted a peak efficiency of $\sim$50\% and
a very narrow angular scan (25\,$\mu$rad FWHM).

%                 SPS U angular scans  MC & Exp

Another SPS experiment  employed a crystal with an
amorphous layer at the edge to suppress the channeling
in the first passage of the protons \cite{sps}.
The extraction efficiency with this crystal was indeed
of the same order of magnitude as found without an amorphous layer,
thus confirming
the theoretical prediction \cite{bi78} that the first-pass channeling is
suppressed in the SPS crystals.

In order to understand some overestimate of the peak efficiency
in the model, we made
a more detailed simulation \cite{book}.
Overestimate of the channeling efficiency might mean an underestimate
of the scattering and/or losses in the multipasses in crystal.
It is clear that the parameters influencing  crystal extraction
are not defined perfectly;
there are several unknowns in the model, such as
the impact parameters and quality of the crystal edge.

In the subsequent simulations
the realistic details of the crystal design,
such as the ``legs of U"
(the scattering here was missed previously) were introduced.
The window for the extracted protons was $\pm$30 $\mu$rad
($\pm$2 $\theta_{\rm c}$)
from the extraction line, in order to match the
experimental procedure (earlier, all protons bent at $>$8.0 mrad
were accepted).

Table \ref{tab4} shows the computed peak efficiency as a function
of the septum width $t$ (modelled as an amorphous layer)
of the U-shaped crystal.
The dependence on $t$ is rather weak; this agrees with the
experiment where the 30-$\mu$m amorphous layer did not affect
the efficiency.

These simulations have been repeated with the energies of
14 and 270 GeV, where new measurements have been done at the SPS.
The results are shown in Table 3.

\begin{table}[htb]
\caption{
The peak efficiency $F$ (\%) for different septum widths $t$ ($\mu$m).
The statistical error is 0.6 \%.}
\label {tab4}
\begin{center}
\begin{tabular}{cccccc}
\hline
 & & & & & \\
 $t$ ($\mu$m) &  1  &  20  &  50  &  100  &  200 \\
 & & & & & \\
\hline
 & & & & & \\
 $F$ (\%)     & 13.9 & 12.4 & 12.9 & 10.9 &  8.2  \\
 & & & & & \\
\hline
\end{tabular}
\end{center}
\end{table}

%                             f ( L )
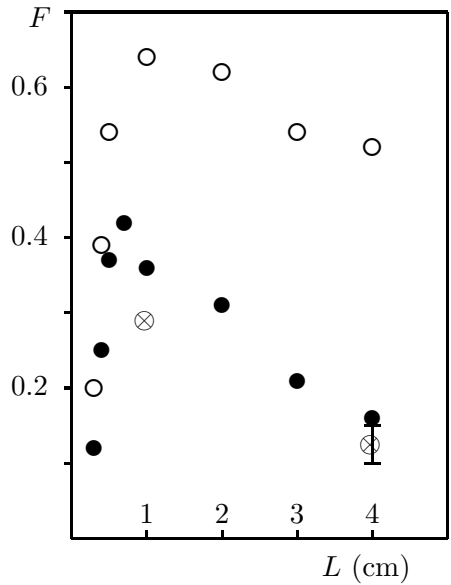
\begin{figure}[htb]
\begin{center}
\setlength{\unitlength}{1.mm}
\begin{picture}(80,69)(-10,-3)
\thicklines

\put(3,20){\circle{2}}
\put(4,39){\circle{2}}
\put(5,54){\circle{2}}
\put(10,64){\circle{2}}
\put(20,62){\circle{2}}
\put(30,54){\circle{2}}
\put(40,52){\circle{2}}

\put(3,12){\circle*{2}}
\put(4,25){\circle*{2}}
\put(5,37){\circle*{2}}
\put(7,42){\circle*{2}}
\put(10,36){\circle*{2}}
\put(20,31){\circle*{2}}
\put(30,21){\circle*{2}}
\put(40,16){\circle*{2}}

\put(38.2,11.4){\small $\otimes$}
\put(8.2,28.0){\small $\otimes$}

\put(0,0) {\line(1,0){50}}
\put(0,0) {\line(0,1){70}}
\put(0,70) {\line(1,0){50}}
\put(50,0){\line(0,1){70}}
\multiput(10,0)(10,0){5}{\line(0,1){1}}
\put(9.5,2.){\makebox(1,.5)[b]{1}}
\put(19.5,2.){\makebox(1,.5)[b]{2}}
\put(29.5,2.){\makebox(1,.5)[b]{3}}
\put(39.5,2.){\makebox(1,.5)[b]{4}}
\multiput(0,10)(0,10){7}{\line(-1,0){1}}
\put(-8,20){\makebox(1,.5)[l]{0.2}}
\put(-8,40){\makebox(1,.5)[l]{0.4}}
\put(-8,60){\makebox(1,.5)[l]{0.6}}

\put(-7,68){ $F$}
\put(32,-5){ $L$ (cm)}
\put(40,10){\line(0,1){5}}
\put(39,15){\line(1,0){2}}
\put(39,10){\line(1,0){2}}

\end{picture}
\end{center}
\caption{
The SPS extraction efficiency vs crystal length.
For a perfect surface (o) and
septum width $t$=1 $\mu$m ($\bullet$).
The $\otimes$ are for the U-shaped design and $t$=20 $\mu$m.
Also shown is the measured range of efficiencies, 10--15\%
for the 4-cm U-shaped crystal.
}	\label{fl}
\end{figure}

The length of the Si crystal used in the experiment
is optimal to bend the 120\,GeV proton beam by 8.5\,mrad
with a {\em single} pass.
The efficiency  of the {\em multi}-pass extraction
is defined by the processes
of channeling, scattering, and nuclear interaction in the crystal,
which depend essentially on the crystal length $L$.
As the scattering is added,
it is qualitatively obvious that
the optimal length is reduced as compared to bending with a single pass.

The optimization with the simulations was made
with the assumption of a uniform crystal curvature, Fig.\,\ref{fl}.
For a perfect surface there is almost no dependence
for $L \ge $ 1\,cm in the range studied,
but for an imperfect surface there is an
important dependence. A new optimum around
$L \simeq $ 0.7\,cm
almost doubles the efficiency as compared to
that for the 3\,cm crystal.
Figure\,\ref{fl} shows also two points from a simulation
 with a U-shaped design and $t$=20 $\mu$m.
The shorter crystal had 1-mm ``legs" and 8-mm bent part (10 mm in total),
and has shown an efficiency near 30 \%.

\section{The Tevatron Experiment}

The Tevatron extraction experiment has provided
another check of theory at a substantially higher energy
of 900 GeV.
A detailed report of predictions for this experiment from the Monte Carlo
simulations was published in Ref. \cite{pre3},
and the experimental data can be found in
\cite{e853}.

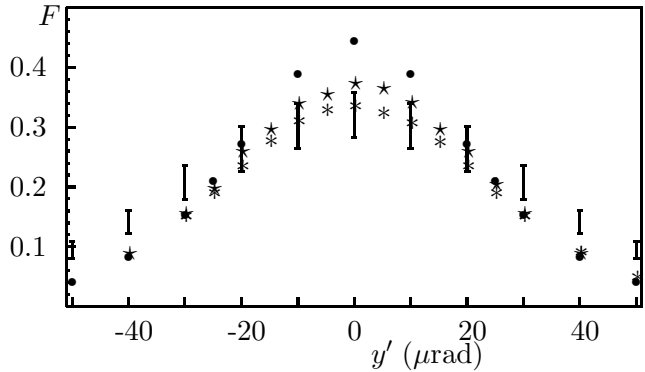
\begin{figure}[bth]
\begin{center}
\setlength{\unitlength}{0.75mm}\thicklines
\begin{picture}(100,55)(-50,-4)

\put(-51,0) {\line(1,0){102}}
\put(-51,0) {\line(0,1){53}}
\put(-51,53) {\line(1,0){102}}
\put( 51,0) {\line(0,1){53}}
\multiput(-50,0)(10,0){11}{\line(0,1){1.5}}
\multiput(-51,10.6)(0,10.6){4}{\line(1,0){1.5}}
\multiput(-51,0)(0,2.12){25}{\line(1,0){.5}}
\put(-40,-6){\makebox(2,1)[b]{-40}}
\put(-20,-6){\makebox(2,1)[b]{-20}}
\put(0,-6){\makebox(.2,.1)[b]{0}}
\put(20,-6){\makebox(2,1)[b]{20}}
\put(40,-6){\makebox(2,1)[b]{40}}
\put(-61,10.6){\makebox(.2,.1)[l]{0.1}}
\put(-61,21.2){\makebox(.2,.1)[l]{0.2}}
\put(-61,31.8){\makebox(.2,.1)[l]{0.3}}
\put(-61,42.4){\makebox(.2,.1)[l]{0.4}}

\put(-56,50){$F$}
\put(3,-10){$y'\; (\mu$rad)}

\put(0,30){\line(0,1){8}}
\put(-0.5,38){\line(1,0){1}}
\put(-0.5,30){\line(1,0){1}}
\put(10,28){\line(0,1){8}}
\put(9.5,28){\line(1,0){1}}
\put(9.5,36){\line(1,0){1}}
\put(20,24){\line(0,1){8}}
\put(19.5,24){\line(1,0){1}}
\put(19.5,32){\line(1,0){1}}
\put(30,19){\line(0,1){6}}
\put(29.5,19){\line(1,0){1}}
\put(29.5,25){\line(1,0){1}}
\put(40,13){\line(0,1){4}}
\put(39.5,13){\line(1,0){1}}
\put(39.5,17){\line(1,0){1}}
\put(50,8.5){\line(0,1){3}}
\put(49.5,8.5){\line(1,0){1}}
\put(49.5,11.5){\line(1,0){1}}
\put(-10,28){\line(0,1){8}}
\put(-10.5,28){\line(1,0){1}}
\put(-10.5,36){\line(1,0){1}}
\put(-20,24){\line(0,1){8}}
\put(-20.5,24){\line(1,0){1}}
\put(-20.5,32){\line(1,0){1}}
\put(-30,19){\line(0,1){6}}
\put(-30.5,19){\line(1,0){1}}
\put(-30.5,25){\line(1,0){1}}
\put(-40,13){\line(0,1){4}}
\put(-40.5,13){\line(1,0){1}}
\put(-40.5,17){\line(1,0){1}}
\put(-50,8.5){\line(0,1){3}}
\put(-50.5,8.5){\line(1,0){1}}
\put(-50.5,11.5){\line(1,0){1}}

\put(10,41.2){\circle*{1.5}}
\put(20,28.8){\circle*{1.5}}
\put(25,22.2){\circle*{1.5}}
\put(30,16.2){\circle*{1.5}}
\put(40,8.7){\circle*{1.5}}
\put(50,4.3){\circle*{1.5}}
\put(0,47){\circle*{1.5}}
\put(-10,41.2){\circle*{1.5}}
\put(-20,28.8){\circle*{1.5}}
\put(-25,22.2){\circle*{1.5}}
\put(-30,16.2){\circle*{1.5}}
\put(-40,8.7){\circle*{1.5}}
\put(-50,4.3){\circle*{1.5}}

\put(-1,34.3){$\ast$}
\put(-6,33.6){$\ast$}
\put(-11,31.6){$\ast$}
\put(-16,28.1){$\ast$}
\put(-21,23.7){$\ast$}
\put(-26,19){$\ast$}
\put(-31,14.8){$\ast$}
\put(4,33.1){$\ast$}
\put(9,31.3){$\ast$}
\put(14,28){$\ast$}
\put(19,23.7){$\ast$}
\put(24,18.9){$\ast$}
\put(29,14.8){$\ast$}
\put(39,8.4){$\ast$}
\put(49,4){$\ast$}

\put(-1,38.4){$\star$}
\put(-6,36.4){$\star$}
\put(4,37.6){$\star$}
\put(-11,34.9){$\star$}
\put(-16,30.2){$\star$}
\put(-21,26.4){$\star$}
\put(-26,19.8){$\star$}
\put(-31,15.3){$\star$}
\put(-41,8.2){$\star$}
\put(9,35.1){$\star$}
\put(14,30.2){$\star$}
\put(19,26.4){$\star$}
\put(24,20.5){$\star$}
\put(29,15.3){$\star$}
\put(39,8.2){$\star$}

\end{picture}
\end{center}
\caption {
Vertical angular scan of the overall efficiency
for the perfect horizontal alignment, $x'$=0.
Ideal crystal ($\bullet$);
imperfect crystal: ($\star$)
with $t$=1 $\mu$m, ($\ast$) is the same with $t$=50 $\mu$m.
Also shown is the measured peak efficiency and angular scan.
      }	\label{r1}
\end{figure}

In our computer model we have investigated three options:
a crystal with ideal surface, one with a septum width
(amorphous layer) of $t$=1 $\mu$m, and one with $t$=50 $\mu$m.
The  crystal bending shape and other details were as used
later in the experiment.
Figure \ref{r1} from Ref.{pre3} shows that there is little difference
between the three options; the peak efficiency is about 35-40\%,
and the angular scan FWHM is 50-55 $\mu$rad.
This insensitivity to the crystal surface quality is due
to the set-up different from that used in other experiments;
as a result, the starting divergence of incident protons
at the crystal was not small and hence less sensitive to edge scattering.

The measured peak efficiency was about 30\%.
This value, together with the measured angular scan,
is superimposed in Figure \ref{r1} on the theoretical expectation,
showing a rather good agreement.

%                   SCAN over Y'

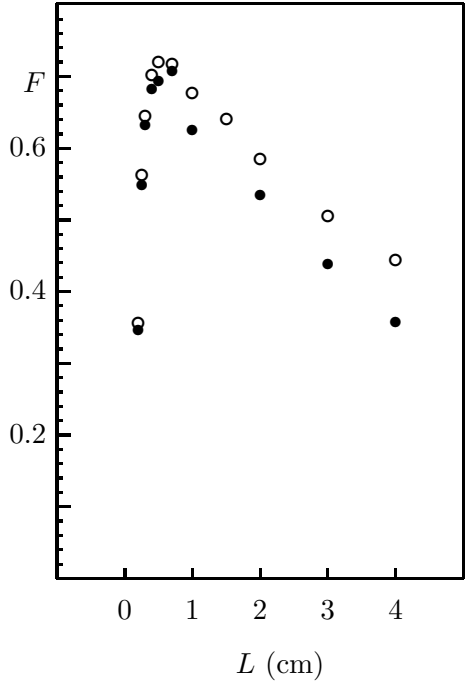
\begin{figure}[htb]
\begin{center}
\setlength{\unitlength}{.9mm}
\begin{picture}(80,103)(-20,-15)
\thicklines
%\Large

\put(2,37.7){\circle{1.5}}
\put(2.5,59.6){\circle{1.5}}
\put(3,68.4){\circle{1.5}}
\put(4,74.5){\circle{1.5}}
\put(5,76.4){\circle{1.5}}
\put(7,76.0){\circle{1.5}}
\put(10,71.7){\circle{1.5}}
\put(15,67.9){\circle{1.5}}
\put(20,62.){\circle{1.5}}
\put(30,53.6){\circle{1.5}}
\put(40,47.0){\circle{1.5}}

\put(2,36.7){\circle*{1.5}}
\put(2.5,58.1){\circle*{1.5}}
\put(3,67.){\circle*{1.5}}
\put(4,72.4){\circle*{1.5}}
\put(5,73.6){\circle*{1.5}}
\put(7,75.0){\circle*{1.5}}
\put(10,66.3){\circle*{1.5}}
\put(20,56.7){\circle*{1.5}}
\put(30,46.5){\circle*{1.5}}
\put(40,38){\circle*{1.5}}

\put(-10,0) {\line(1,0){60}}
\put(-10,0) {\line(0,1){85}}
\put(-10,85) {\line(1,0){60}}
\put(50,0){\line(0,1){85}}
\multiput(0,0)(10,0){6}{\line(0,1){1.5}}
\put(-.5,-6.){\makebox(1,.5)[b]{0}}
\put(9.5,-6.){\makebox(1,.5)[b]{1}}
\put(19.5,-6.){\makebox(1,.5)[b]{2}}
\put(29.5,-6.){\makebox(1,.5)[b]{3}}
\put(39.5,-6.){\makebox(1,.5)[b]{4}}
\multiput(-10,0)(0,10.6){8}{\line(1,0){2}}
\multiput(-10,0)(0,2.12){40}{\line(1,0){.75}}
\put(-17,21.2){\makebox(1,.5)[l]{0.2}}
\put(-17,42.4){\makebox(1,.5)[l]{0.4}}
\put(-17,63.6){\makebox(1,.5)[l]{0.6}}

\put(-15,72){$F$}
\put(15,-14){ $L$ (cm)}

\end{picture}
\caption {
Efficiency as a function of $L$
for the ideal (o) and imperfect ($\bullet$), $t$=1 $\mu$m, crystals.
      }	\label{fl2}
\end{center}
\end{figure}

The efficiency of extraction can again be increased with the use
of a shorter crystal.
Fig. \ref{fl2} shows the extraction efficiency dependence
on the crystal length $L$, for uniform bending of crystal.
The efficiency is maximal, near 70 \%,
in the length range from 0.4 to 1.0 cm.

%                             f ( L )

\section{Analytical Theory of Multipass
  Crystal Extraction}

An analytical theory of multipass crystal extraction
would be highly helpful in understanding the
experimental results.
Below we describe a simple theory for the
extraction efficiency \cite{theory}.

Suppose that a beam with divergence $\sigma$, Gaussian distribution,
is aligned to the crystal planes. Then as many as
\begin{equation}
(2\theta_c/\sqrt{2\pi}\sigma ) (\pi x_c/2d_p)
\end{equation}
particles get channeled in the initial straight part of the crystal.
Here $\theta_c$ stands for the critical angle of channeling,
$d_p$ the interplanar spacing, $x_c\approx d_p/2-a_{TF}$
the critical distance, $a_{TF}$ being the Thomas-Fermi screening
distance.

We shall first consider the case
where particles first come to
the crystal with nearly zero divergence, due to very small impact
parameters.
We assume then that any particle always crosses the full crystal
length; that pass 1 is like through an amorphous matter but any
further pass is like through a crystalline matter;
that there are no aperture restrictions; and that the particles
interact only with the crystal not a holder.
After some turns in the accelerator ring, the scattered particles come
to the crystal with rms divergence as defined by scattering in the
first pass:
\begin{equation}
\sigma_1 = (E_s/pv)(L/L_R)^{1/2} ,
\end{equation}
where $E_s$=13.6 MeV, $L$ is the crystal length, $L_R$ the radiation length,
$pv$ the particle momentum times velocity.

After $k$ passes the divergence is $\sigma_k=k^{1/2}\sigma_1$.
The number of particles lost in nuclear interactions is
1$-\exp(-kL/L_N)$ after $k$ passes; $L_N$ is the interaction length.
In what follows we shall first assume that
 the crystal extraction efficiency is substantially smaller than
100 \% (which has actually been the case so far), i.e. the circulating
particles are removed from the ring predominantly through the
nuclear interactions, not through channeling.

That pulled together, we obtain the multipass channeling
efficiency by summation over $k$ passes, from 1 to infinity:
\begin{equation}
F_C=\left(\frac{\pi}{2}\right)^{1/2}\frac{\theta_cx_c}{\sigma_1d_p}
   \times \Sigma(L/L_N)
\end{equation}
where
\begin{equation}
\Sigma(L/L_N)= \Sigma_{k=1}^{\infty} k^{-1/2}\exp(-kL/L_N)
\end{equation}
may be called a "multiplicity factor" as it just tells how much
the single-pass efficiency is amplified in multipasses.

A fraction $1-T$ of channeled particles is to be lost along the bent
crystal due to scattering processes and centripetal effects.
Then the multipass extraction efficiency is
\begin{equation}
F_E=F_C\times T=
  \left(\frac{\pi}{2}\right)^{1/2}\frac{\theta_cx_c}{\sigma_1d_p}
    \times \Sigma(L/L_N) \times T
\end{equation}
We shall use an analytical approximation (as used also in \cite{a0})
for silicon
\begin{equation}
  T = (1-p/3R)^2 \exp\left(-\frac{L}{L_d(1-p/3R)^2}\right),
\end{equation}
where $p$ is in GeV/c, and $R$ is in cm;
$L_d$ is dechanneling length for a straight crystal.
The first factor in $T$ describes a centripetal dechanneling.
E.g., at $pv/R$=0.75 GeV/cm (which is close to the highest
values used in extraction)
our approximation gives $(1-p/3R)^2$=0.563
whereas Forster et al.\cite{fors} measured 0.568$\pm$0.027.
We shall use the theoretical formula for $L_d$
\cite{book}.
The sum (4)
can be approximated as follows:
\begin{equation}
\Sigma(L/L_N) \simeq (\pi L_N/L)^{1/2}-1.5
\end{equation}

Let us check the theory, first against the CERN SPS data \cite{pac97}
where the crystal
extraction efficiency was measured at 14, 120, and 270 GeV
(Table 1).
\begin{table}[htb]
\begin{center}
\caption{Extraction efficiencies (\%) from the SPS experiment,
 theory, and detailed simulations.}
\begin{tabular}{cccc}
 & & & \\
                $pv$(GeV) &    SPS     & Theory & Monte Carlo \\
\hline
 & & & \\
                14 &   0.55$\pm$0.30&  0.30 & 0.35$\pm$0.07 \\
               120 &   15.1$\pm$1.2 & 13.5 & 13.9$\pm$0.6  \\
               270 &   18.6$\pm$2.7 & 17.6 & 17.8$\pm$0.6
\end{tabular}
\end{center}
\end{table}

The Tevatron extraction experiment at 900 GeV provides
another check.
Here a slight modification of the formulas is needed to
account for the non-zero starting divergence, namely
$\sigma_0$=11.5 $\mu$rad (rms).
This results in the change in Eq.(4):
\begin{equation}
\Sigma(L/L_N)= \Sigma_{k=1}^{\infty}
  (k+\sigma_0^2/\sigma_1^2)^{-1/2}\exp(-kL/L_N)
\end{equation}
Since in this experiment Si(111) planes were used, consisting
of narrow (1/4 weight) and wide (3/4 weight) channels,
this is to be taken into account in Eq.(5).
Eq.(5) then gives an extraction efficiency of 40.8~\%.
However, a minor correction to the theoretical value
is discussed below.

As the extraction efficiency is getting high,
our earlier assumption that the nuclear interactions
dominate over the crystal channeling may need correction.
To take into account the fact that the circulating particles are efficiently
removed from the ring by a crystal extraction as well,
one would require a {\em recurrent} procedure of summation:
instead of $\Sigma F_k$ one has to sum $\Sigma F^*_k$,
where $F^*_k$=$F_k(1-F^*_{k-1})$.
This ``recurrent'' correction doesn't affect our earlier
SPS calculation at 14 GeV and makes $\sim$1\% drop to the efficiencies
at 120 and 270 GeV listed in Table 1.
For Tevatron this correction constitutes $-$6.7\%,
converting 40.8\% into 34.1\%,
more into line with the measurement.

To see the dependence of extraction efficiency on the
microscopic properties of the crystal material
and on the particle energy,
let us use the well-known theoretical expressions for
$\theta_c$=(4$\pi Nd_pZe^2a_{TF}/pv)^{1/2}$,
radiation length
$L_R$=137/$[4Z(Z+1)r_e^2N\ln(183Z^{-1/3})]$,
and $E_s$=2$\sqrt{2\times 137}m_ec^2$,
where $N$ is the number of atoms per unit volume of crystal.
The multipass extraction efficiency is then
\begin{equation}
F_E= \frac{\pi}{4}\left(\frac{x_c^2a_{TF}}{L(Z+1)d_pr_e\ln(183Z^{-1/3})}
\right)^{1/2}
\end{equation}
  $$\times \left(\frac{pv}{m_ec^2}\right)^{1/2} T \Sigma(L/L_N)$$
here $m_e$ is the electron mass, $r_e$ the classical electron radius.
Despite of the simplifications done, this equation still
predicts the SPS efficiency of 15.7\% at 120 GeV which is
within the experimental error limits.

\begin{figure}[htb]
\begin{center}
\setlength{\unitlength}{.63mm}
\begin{picture}(110,90)(0,-10)
\thicklines
\linethickness{.25mm}
\put(    40.,12.8)  {\circle*{3}}
\put(    30.,17.9)  {\circle*{3}}
\put(    20.,25.)  {\circle*{3}}
\put(    10.,26.6)  {\circle*{3}}
\put(    5.,0.)  {\circle*{3}}
\put(    8.,20.)  {\circle*{3}}
\put(    6.,4.6)  {\circle*{3}}
\put(    50.,9.3)  {\circle*{3}}
\put(    60.,6.9)  {\circle*{3}}
\put(    70.,5.2)  {\circle*{3}}
\put(    80.,3.9)  {\circle*{3}}

\put(    40.,39)  {\circle {3}}
\put(    30.,48)  {\circle {3}}
\put(    20.,58)  {\circle {3}}
\put(    10.,67)  {\circle {3}}
\put(    5.,61)  {\circle {3}}
\put(    8.,67)  {\circle {3}}
\put(    4.,46)  {\circle {3}}
\put(    50.,33)  {\circle {3}}
\put(    60.,28)  {\circle {3}}
\put(    70.,24)  {\circle {3}}
\put(    80.,21)  {\circle {3}}

\put(    40.,39)  {$\star$}
\put(    30.,21)  {$\star$}
\put(    20.,2.)  {$\star$}
\put(    50.,45)  {$\star$}
\put(    60.,46)  {$\star$}
\put(    70.,44)  {$\star$}
\put(    80.,42)  {$\star$}
\put(    90.,40)  {$\star$}

%\put(40,13.9) {\line(0,1){2.4}}
%\put(39,15.1) {\line(1,0){2}}

\put(0,0) {\line(1,0){100}}
\put(0,0) {\line(0,1){90}}
\put(0,90) {\line(1,0){100}}
\put(100,0){\line(0,1){90}}
\multiput(10,0)(10,0){9}{\line(0,-1){2}}
\multiput(0,10)(0,10){8}{\line(1,0){2}}
\multiput(0,5)(0,5){16}{\line(1,0){1.4}}
\put(10,-7){\makebox(1,1)[b]{1}}
\put(20,-7){\makebox(1,1)[b]{2}}
\put(30,-7){\makebox(1,1)[b]{3}}
\put(40,-7){\makebox(1,1)[b]{4}}
\put(50,-7){\makebox(1,1)[b]{5}}
\put(60,-7){\makebox(1,1)[b]{6}}
\put(70,-7){\makebox(1,1)[b]{7}}
\put(80,-7){\makebox(1,1)[b]{8}}
\put(90,-7){\makebox(1,1)[b]{9}}
\put(-15,20){\makebox(1,.5)[l]{0.2}}
\put(-15,40){\makebox(1,.5)[l]{0.4}}
\put(-15,60){\makebox(1,.5)[l]{0.6}}
\put(-15,80){\makebox(1,.5)[l]{0.8}}

\put(-17,50){\large F}
\put(73,-14){\large $L$ (cm)}

\end{picture}
\end{center}
\caption{ The extraction efficiency, Eq.(5), as a function of
the crystal length $L$;  for the SPS ($\bullet$), Tevatron (o),
and Large Hadron Collider ($\star$).
}
  \label{a3}
\end{figure}
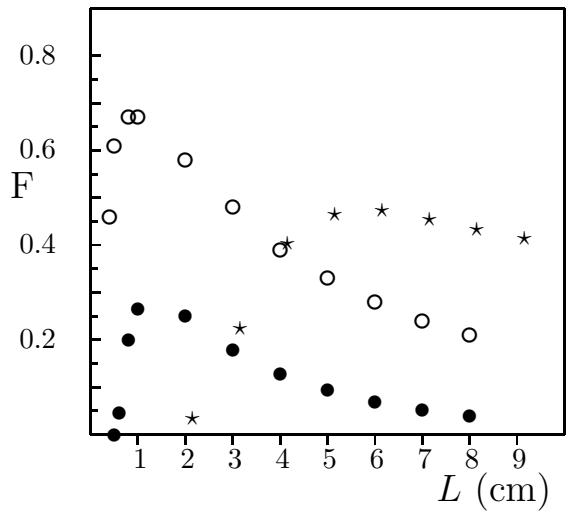

Figure \ref{a3} shows the $F_E(L)$ dependence for extraction at the
120-GeV SPS, 900-GeV Tevatron, and 7-TeV Large Hadron Collider
(where 0.7 mrad deflection angle is assumed);
in all the cases the crystal bent part was
0.75 of the full length.
One can see that the analytical dependences $F_E(L)$ are very close
to those obtained earlier in Monte Carlo simulations \cite{nimb}.
The same maxima at the same optimal lengths are predicted.

Formula (5) predicts a high efficiency of multipass
extraction at a multi-TeV LHC, about 45 \%, with the
optimal length of Si(110) crystal being 6$\pm$1 cm.

%\end{document}

\section{IHEP Experiment}

The pioneering crystal extraction experiments at Protvino
IHEP 70-GeV accelerator were made \cite{ass1}
before any computer simulations of this kind.
This is why we prefer to mention a new IHEP experiment
planned for November 1997 where one could make predictions
in advance.

\begin{figure}[htb]
\begin{center}
\setlength{\unitlength}{.5mm}
\begin{picture}(110,60)(10,-3)
\thicklines
\linethickness{.3mm}
\put(    140.,2.8)  {\circle*{3}}
\put(    125.,6.0)  {\circle*{3}}
\put(    115.,9.2)  {\circle*{3}}
\put(    105.,14.6)  {\circle*{3}}
\put(    95.,21.6)  {\circle*{3}}
\put(    85.,29.4)  {\circle*{3}}
\put(    75.,37.)  {\circle*{3}}
\put(    10.,2.6)  {\circle*{3}}
\put(    25.,7.2)  {\circle*{3}}
\put(    35.,13.6)  {\circle*{3}}
\put(    45.,25.8)  {\circle*{3}}
\put(    55.,38.4)  {\circle*{3}}
\put(    65.,42.8)  {\circle*{3}}

\put(    115.,2.8)  {\circle{3}}
\put(    105.,5.8)  {\circle{3}}
\put(    95.,9.4)  {\circle{3}}
\put(    85.,12.4)  {\circle{3}}
\put(    75.,16.6)  {\circle{3}}
\put(    5.,2.2)  {\circle{3}}
\put(    15.,3.8)  {\circle{3}}
\put(    35.,11.6)  {\circle{3}}
\put(    45.,13.2)  {\circle{3}}
\put(    55.,16.6)  {\circle{3}}
\put(    65.,18.8)  {\circle{3}}

\put(0,0) {\line(1,0){140}}
\put(0,0) {\line(0,1){60}}
\put(0,60) {\line(1,0){140}}
\put(140,0){\line(0,1){60}}
\multiput(15,0)(10,0){13}{\line(0,-1){2}}
\multiput(0,20)(0,20){3}{\line(1,0){2}}
\multiput(0,4)(0,4){15}{\line(1,0){1.4}}
\put(13,-7){\makebox(1,1)[b]{--0.5}}
\put(64,-7){\makebox(1,1)[b]{0}}
\put(114,-7){\makebox(1,1)[b]{0.5}}
\put(-15,20){\makebox(1,.5)[l]{0.1}}
\put(-15,40){\makebox(1,.5)[l]{0.2}}

\put(-17,50){ F}
\put(73,-14){ Angle (mrad)}

\end{picture}
\end{center}
\caption{
The angular scan of the extraction efficiency
as seen in Monte Carlo simulations for 70-GeV IHEP experiment.
Crystal without targets ($\bullet$), and with Be target (o).
}
  \label{i1}
\end{figure}
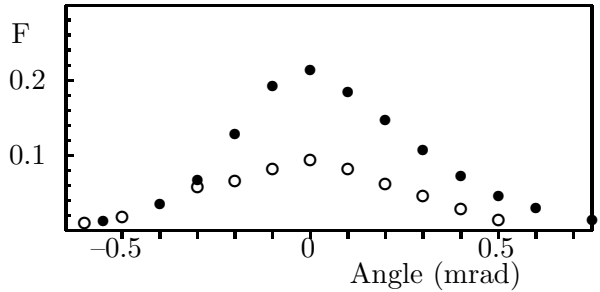

This experiment employs a very short (7 mm along the beam) silicon
crystal bent a small angle of 1.75 mrad.
Figure \ref{i1} shows the angular scan of the extraction efficiency
as seen in Monte Carlo simulations.
The peak efficiency is rather modest, about 20\%,
because of a big effective divergence of the protons
at crystal w.r.t. the crystal planes
(part of it is due to the crystal design,
another part is due to the beam phase space geometry).

As the experiment would also investigate a co-existence
of crystal extraction with simultaneous work of two internal
targets, this option was simulated as well.
We have seen practically no influence on the crystal efficiency
from a very thin carbon target, whereas a 3-cm long beryllium
target could decrease the extraction efficiency
(defined as the ratio of protons extracted to protons lost
in nuclear interactions in the crystal) up to factor of two.
Figure \ref{i1} shows the angular scan in this case also.

\section{Future Applications}

The progress in crystal extraction studies at CERN and Fermilab
has been stimulated by the prospects of application of this technique
for extraction of a parasitic beam from a large hadron collider
for a fixed target physics.
Such an extraction is quite feasible from the standpoint of channeling
physics. The theory and simulations predict the extraction efficiency
of about 50\% even under the most conservative assumptions
on the crystal design and edge quality.

Another discussed option is extraction from the Tevatron \cite{a0}
with required minimal angle of 16.4 mrad. In our simulations
of this option with use of the same set-up as in the E853 experiment,
the efficiency is expected to be 6.3$\pm$0.7\% with Si crystal
of $\sim$12 cm length even if the first-pass channeling is fully suppressed.
However, if channeling in the first encounter is efficient
(good crystal edge), the efficiency becomes as high as 23\% with the
use of optimal 5-cm long Ge(110) crystal.
Notice, that this figure---over 20\% efficiency of bending at 16.4 mrad
by a Ge(110) crystal---is already demonstrated experimentally at CERN
with a 200 GeV beam \cite{ge}!

One very interesting option is a crystal use in the beam collimation systems.
A principle problem for an amorphous collimator is the edge scattering
causing a leak of particles incident closer than $\sim$1 $\mu$m
to the collimator edge. Furthermore, if collimator of length $L$
is misaligned by an angle $\theta$, the inefficient edge thickness
is increased by $L\theta$; therefore, an amorphous collimator
should be aligned with accuracy of order $\theta\ll$1 $\mu$m/$L$
$\simeq$2 $\mu$rad (for $L$=450 mm\cite{white})!
Compare this with critical angles for crystals---order of 20 $\mu$rad
at 100 GeV and order of 2 $\mu$rad at 7 TeV.
Of course, it is much easier to align crystals than huge collimators.

An edge leak doesn't exist in crystalline material for channeled particles.
The simplest idea is to put a bent crystal in front and at the edge
of a heavy collimator.
A large fraction of incident particles is bent by the crystal
some small angle of 0.1-0.3 mrad toward the depth of the collimator,
and hence fully absorbed
(this idea has something common with the idea of a magnetized
collimator\cite{white}).
The collimator has only to deal with the remaining particles,
unchanneled in crystal.
According to our Monte Carlo simulations, the efficiency of bending
of a parallel beam is about 90\% for a 1-TeV beam and 2-cm long Si(110)
crystal bent 0.2 mrad, and for a 7-TeV beam and 5-cm long Si(110) crystal
bent 0.1 mrad. Notice, that the experimentally demonstrated
record of bending efficiency at CERN is already 60\% for 2 mrad
bending angle at the energy of 0.45 TeV \cite{ge}!
Hence, under the optimal conditions the inefficiency the collimation system
can be reduced by factor of 10. If it were a two-stage collimation system,
and both stages equipped by bent crystals, the inefficiency of the
whole system would be reduced by factor of 100.

Notice that this simplest idea doesn't affect the optics of the collimation
design. One could take the existing collimation system
and just add crystals to improve its efficiency.

More advanced idea would be to separate a bent crystal from
a heavy collimator, and to optimize its position w.r.t.
the collimator. This idea has been discussed and simulated
in Ref.\cite{mas}.

\section{Acknowledgements}

The author is much indebted to Dick Carrigan, Nikolai Mokhov,
and Thornton Murphy for kind hospitality and support
during the author's stay at Fermilab.

\end{document}